# An interpretable planning bot for pancreas stereotactic body radiation therapy

Jiahan Zhang[1]*, Chunhao Wang[1], Yang Sheng[1], Manisha Palta[1], Brian Czito[1], Christopher Willett[1], Jiang Zhang[1], P James Jensen[1], Fang-Fang Yin[1], Qiuwen Wu[1], Yaorong Ge[2], Q Jackie Wu[1]

1. Department of Radiation Oncology, Duke University Medical Center, Durham NC, 27705

2. University of North Carolina at Charlotte, Charlotte NC, 28223

## Abstract

**Purpose**

Pancreas stereotactic body radiotherapy (SBRT) treatment planning requires planners to make sequential, time consuming interactions with the treatment planning system (TPS) to reach the optimal dose distribution. We seek to develop a reinforcement learning (RL)-based planning bot to systematically address complex tradeoffs and achieve high plan quality consistently and efficiently.

**Methods**

The focus of pancreas SBRT planning is finding a balance between organs-at-risk (OAR) sparing and planning target volume (PTV) coverage. Planners evaluate dose distributions and make planning adjustments to optimize PTV coverage while adhering to OAR dose constraints. We have formulated such interactions between the planner and the TPS into a finite-horizon RL model. First, planning status features are evaluated based on human planners' experience and defined as planning states. Second, planning actions are defined to represent steps that planners would commonly implement to address different planning needs. Finally, we have derived a "reward" system based on an objective function guided by physician-assigned constraints. The planning bot trained itself with 48 plans augmented from 16 previously treated patients and generated plans for 24 cases in a separate validation set.

**Results**

All 24 bot-generated plans achieve similar PTV coverages compared to clinical plans while satisfying all clinical planning constraints. Moreover, the knowledge learned by the bot can be visualized and interpreted as consistent with human planning knowledge, and the knowledge maps learned in separate training sessions are consistent, indicating reproducibility of the learning process.

**Conclusion**

We have developed a planning bot that generates high quality treatment plans for pancreas SBRT. We have demonstrated that the training phase of the bot is tractable and reproducible, and the knowledge acquired is interpretable. As a result, the RL planning bot can potentially be incorporated into the clinical workflow and reduce planning inefficiencies.

# 1. Introduction

For locally advanced pancreatic cancer patients, one standard of care is concurrent chemotherapy with conventionally fractionated radiation therapy. Due to improvements in motion management, imaging technology, and treatment delivery accuracy, it is now possible to utilize stereotactic body radiotherapy (SBRT) for pancreatic cancer treatment with low risks of radiation-induced toxicity [1]. With SBRT, radiation dose is delivered to patients over shorter periods and without significant delays in systemic therapy [2]. National database studies have suggested that chemotherapy followed by SBRT results in better outcomes than chemotherapy alone or chemotherapy concurrent with conventionally-fractionated intensity-modulated radiotherapy (IMRT) [3,4]. However, the treatment planning of pancreas SBRT poses a challenge to planners given pancreas SBRT treatments are inherently difficult to plan, considering patient-specific planning target volume (PTV) coverage and organs-at-risk (OAR) sparing tradeoff requirements and high inter-patient anatomy variability.

Treatment planning, especially pancreas SBRT planning, is inherently iterative and interactive. The planning process starts with a planner setting initial optimization constraints to the PTV and OARs and executing the optimization algorithm embedded in the treatment planning system (TPS). The initial optimization constraint set will not generate the optimal plan, due to individual anatomy variations. Therefore, the planner is required to iteratively adjust the optimization objectives to make it clinically optimal. Due to the toxicity concerns of the GI structures and their proximity to the PTVs, planners usually rely on a trial-and-error approach and repetitively interact with the TPS to achieve clinical optimality. This process is time-consuming, and the resultant plan quality is highly subjective to planner experience.

Reinforcement learning [5,6] presents a potential solution to this problem. A reinforcement learning agent—in our case, a planning bot—gains decision-making knowledge by repetitively interacting with the surrounding environment (TPS) and evaluating rewards (improvement of the plan dose distribution) associated with the action (changing of optimization objectives). State-action-reward-state-action (SARSA) [7], also known as connectionist Q-learning, is a widely-used reinforcement learning algorithm and has been proven to perform well in wide-ranging real-world applications such as controlling power systems [8], advanced robotics [9], and playing video games [10,11]. It is an efficient, sampling-based algorithm that sequentially changes the knowledge of the agent based on the interactive training process. We have developed a SARSA-based treatment planning bot that assists planners to efficiently achieve consistent and high-quality plans for pancreas SBRT treatments. We hypothesize that, through repetitive interactions with the TPS, the autonomous planning bot can learn to make appropriate adjustments given anatomical information and intermediate planning results, and ultimately design clinical optimal plans.

# 2. Methods and Materials

Pancreas SBRT treatment planning is a highly interactive process. Although the TPS can optimize plans with respect to the objective function given by the planner, the setting of planning objectives is highly dependent on the shape, size, and location of the PTVs. The planner usually interacts with the TPS multiple times and performs various actions including adjusting dose-volume constraints and creating necessary auxiliary structures in order to get desirable dose distributions. The action-making decisions are guided by the current planning status and the planner's prior experience-based assessment.

Here we adopt a SARSA reinforcement-learning framework to perform these tasks systematically. The formulation of SARSA is as follows [7]:

$$Q(s,a) \leftarrow Q(s,a) + \alpha \cdot [r(s,a,s') + \gamma \cdot Q(s',a') - Q(s,a)],$$

where $s$ and $a$ denote the current state and action; $s'$ and $a'$ denote the next state and action; $Q$ denotes the value function; $r$ denotes the immediate reward; $\alpha$ denotes the learning rate of the bot; and $\gamma$ denotes the discount factor of the system. In particular, the action value function $Q$ predicts the expected long-term reward. The goal of the iteration during the training phase is to parameterize $Q$, which can be subsequently used to guide future decision making. With linear function approximation, we formulate the action value function of the treatment planning RL problem as:

$$Q_\theta(s,a) = \theta^T \varphi(s,a),$$

where $Q_\theta(s,a)$ represents the expected final score value at state $s$ when action $a$ is taken, $\theta^T$ denotes the feature vector that will be learned through the training process, and $\varphi(s,a)$ is a set of features carefully engineered to reduce the complexity of the reinforcement learning problem without losing out on generalization. In our implementation, the feature $\varphi(s,a)$ is generated as an outer product of a state vector $f(s)$ and an action vector $g(a)$: $\varphi(s,a) = vec[\,f(s) \otimes g(a)\,]$. Here $\otimes$ denotes the outer product operator, which multiplies each element of the row vector $f(s)$ to each element of the column vector $g(a)$. The state vector $f(s)$ is formulated as $f(s) = [\triangle D_1, \triangle D_2, \dots, \triangle D_N]$, where $\triangle D_n = \overline{D_n} - D_n$, $n \in [1,2,3,\dots,N]$ denotes the differences between the predicted/estimated dose constraints and the actual dose values at the current iteration. The complete state vector implemented for our pancreas SBRT planning module is listed in the supplementary materials.

**Table 1.** Action options for the RL planning program.

| Action index | Structure | Volume | Dose | Priority | Constraint type |
|---|---|---|---|---|---|
| $A_1, A_2, A_3$ | Primary PTV minus overlapping region with GI OARs with 0mm, 4mm, 6mm expansion | 96% | $\overline{D_{pri}}$* | 80 | Lower |
| $A_4, A_5, A_6$ | Boost PTV minus overlapping region with GI OARs with 0mm, 4mm, 6mm expansion | 96% | $\overline{D_{bst}}$* | 80 | Lower |
| $A_7, A_8, A_9$ | Bowel with 2mm, 4mm, 6mm expansion | 0.5 cm³ | $D_{1cc} - 2\,Gy$ | 80 | Upper |
| $A_{10}, A_{11}, A_{12}$ | Duodenum with 2mm, 4mm, 6mm expansion | 0.5 cm³ | $D_{1cc} - 2\,Gy$ | 80 | Upper |
| $A_{13}, A_{14}, A_{15}$ | Stomach with 2mm, 4mm, 6mm expansion | 0.5 cm³ | $D_{1cc} - 2\,Gy$ | 80 | Upper |
| $A_{16}$ | PTV$_{pri}$ minus PTV$_{bst}$ | 20 % | $D_{20\%} - 2\,Gy$ | 50 | Upper |
| $A_{17}$ | Liver | 50 % | $12\,Gy$ | 50 | Upper |
| $A_{18}$ | Kidneys | 30 % | $12\,Gy$ | 50 | Upper |
| $A_{19}$ | Cord | 0 | $20\,Gy$ | 50 | Upper |

*$\overline{D_{pri}}$ and $\overline{D_{bst}}$ denote the prescription levels for the primary PTV and the boost PTV, respectively.

The action vector $g(a) = [1(a = A_1), 1(a = A_2), ..., 1(a = A_M)]^T$ is an array of $M$ indicators that represent indices of $M$ actions. The $M$ action options are designed based on the actions commonly taken by our clinical planners during pancreas SBRT treatment planning. Since we are taking sequential steps, the vector only has one non-zero component at any step during the iterations. In total, 19 actions are designed to ensure the bot has an optimal choice in any given state that may lead to the optimal plan quality. The actions include adding constraints to liver, kidney, cord, and auxiliary structures associated with stomach, duodenum, bowel, primary PTV, and boost PTV. Full descriptions of the actions are listed in Table 1. It is worth noting that the fixed priorities carried by the actions can be viewed as fixed step sizes. The bot takes one action per interaction and is allowed to take repeated actions.

The reward $r$ is assigned as the plan quality score improvement after each step: $r = S' - S$, where $S$ and $S'$ denote the plan quality score before and after taking the current action, respectively. The plan score metric S is set as a weighted combination of various clinical plan quality metrics:

$$S = -\sum_i W_i max(K_i - \overline{K_i}, 0) - \sum_j W_j max(H_j - \overline{H_j}, 0)^2,$$

where $\overline{K_i}, \overline{H_j}$ denote prescribed soft and hard constraints and $K_i, H_j$ denote achieved soft and hard constraint values. In this study, hard constraints refer to the constraints assigned to bowel, duodenum, stomach and cord. Soft constraints are the constraints for liver and kidney. The plan quality score $S$ is re-evaluated each time the bot takes one action. To keep the notation simple, we assign positive values to upper constraints (OAR sparing, PTV hotspot, dose conformity) and negative values to lower constraints (PTV coverage). The weights are selected carefully to reflect clinical plan quality preferences, which were consulted and reviewed with physician co-investigators during the experiment design. The current implementation focuses on getting as much target boost coverage as possible while satisfying GI structure $D_{1cc}$ dose constraints. This strategy is consistent with our current clinical practice preference, as the boost PTV prescription dose is likely to be higher for therapeutic gains. Different weightings of the plan quality scores produce planning bots with different tradeoff preferences, as the bot's perception of expected long term rewards are directly linked to plan quality scores.

**Algorithm 1.** Iteration scheme of the planning bot training phase

---

Initialize the weighting vector $\theta$.
Set exploration-exploitation factor $\varepsilon$, learning rate $\alpha$, discount factor $\gamma$.
For $E_{max}$ epochs
    For M patients
        Initialize plan, set initial constraints based on a template. Optimize plan.
        Run N times
             Take action $a = \arg\max_a Q_\theta(s, a)$ or a random action ($\varepsilon$-greedy).
            Optimize plan.
            Evaluate features $\varphi(s', a')$ and reward r.
            $Q_\theta(s', a') = \theta^T \varphi(s', a')$
            $\delta = r + \gamma Q_\theta(s', a') - Q_\theta(s, a)$
            $\theta \leftarrow \theta + \eta \delta \varphi(s, a)$

---

The iteration scheme for the planning bot training process is given in Algorithm 1. During each iteration in the training process (Fig. 1a), a random number generator produces a number between 0 and 1, and if the number is larger than the predetermined threshold ε, a random action is taken. Otherwise, optimal policy-based actions indicated by the current Q-function are taken. The introduced randomness in the training process allows the bot possibility to explore different/unseen actions and evaluate the values of these actions associated with the current state. This learning approach is known as ε-greedy. It allows the planning bot to explore the action-value space and acquire planning knowledge without being fully confined to prior experience. In this study, ε is set to gradually decrease over time:

$$\varepsilon = max(0.05, 1 - E/E_{max}),$$

where $E$ and $E_{max}$ denote current epoch number and maximum epoch number, respectively. In each epoch, the planning bot practices planning once on each training case. The value of ε decreases linearly as the number of epochs increases and stays equal or greater to 0.05. It is worth noting that the randomness only exists in the training phase. In the validation phase, the planning bot only follows the guidance of the action-value function in every step.

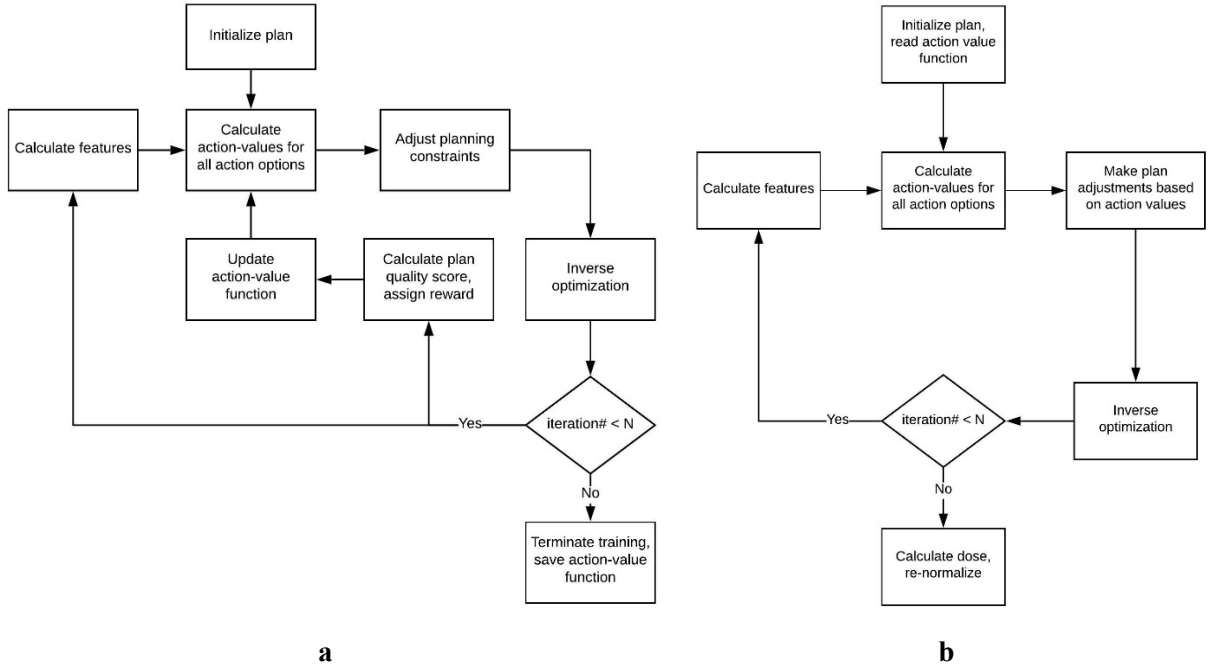

**Figure 1**. The workflow of the proposed RL planning framework: (a) training phase (b) validation/application phase.

The RL training and validation workflow, as shown in Fig. 1, has been implemented in a research TPS environment (Eclipse[TM] Treatment Planning System Version 13.7, Varian Medical Systems, CA). Actions are defined as a set of function-calls inside the TPS during the planning phase, enabled by Eclipse Scripting Application Programing Interface (ESAPI). To evaluate the performance of the proposed planning bot framework, we anonymized and retrieved 40 biopsy-proven pancreatic cancer patients previously treated at our institution. Triphasic imaging technique was used during simulation for target delineation. The primary tumor and adjacent nodal disease were contoured on each imaging sequence. The union of these volumes was used as the internal target volume (ITV) because this volume should provide a good estimation of the tumor motion range during treatments. The boost PTV prescribed to 33 Gy was defined as the GTV with a 2-3 mm margin and minus GI luminal structures. All 40 patients were treated with SIB technique to 25Gy/33Gy in 5 fractions. The OAR constraints of these patients were consistent with the multi-institutional phase II pancreas SBRT study by Herman *et al*[12], in which the authors reported low rates of toxicity.

In this dataset, 22 patients were planned on free-breathing CT, 5 patients were planned on average CT processed from free-breathing 4DCT sequences, and the remaining 13 patients were planned on breath-hold CT. The average sizes of primary PTVs ($PTV_{25Gy}$) and boost PTVs ($PTV_{33Gy}$) were $200 \pm 144$ cm$^3$ and $62 \pm 32$ cm$^3$, respectively. The average volumes of liver and kidneys were $1504 \pm 307$ cm$^3$ and $320 \pm 71$ cm$^3$. From the cohort, 16 patients were randomly selected to train the RL planning bot. We augmented the training set to 48 plans by expanding the $PTV_{33Gy}$ by -2 mm, 0 mm, and 2 mm. The training workflow for a patient, as illustrated in Fig. 1a, consisted of $N=15$ sequential bot-TPS interactions. The RL system was trained with 20 epochs, meaning that the RL bot practiced planning by making 20 different plans for each of the 48 cases in the training set. For each plan, the planning bot initializes with a minimal set of optimization constraints, including PTV lower constraints, kidney, and liver upper constraints. The constraints are given to reduce the number of necessary bot-TPS interactions and accelerate the planning process. The only information carried over from an epoch to another was the weighting vector $\theta$. After the RL bot was fully trained, following the workflow shown in Fig. 1b, we generated treatment plans for the remaining 24 patients and compared them with the clinical treatment plans.

## 3. Results

In order to determine the efficacy of using the proposed RL planning bot in the clinical environment, we validated the plan quality and examined the training process by analyzing the learning behavior of the planning bot, including state specificity of the bot during the training phase, knowledge interpretability, and knowledge reproducibility.

### 3.1. Plan quality and efficiency

It took 5 days to train an RL bot on a single Varian workstation. For the validation set, the bot spent $7.3 \pm 1.0$ min on each case to create a deliverable plan from a set of contours. This is a significant improvement over manual planning, which typically takes 1-2 hours. Figure 2 shows the planning results for cases in the validation set. All OAR constraints have been met by both clinical plans and RL bot plans. All 24 clinical plans and 24 RL plans meet pre-defined GI constraints (V33Gy<1cc). PTV coverages are comparable between RL plans ($PTV_{25Gy}$: $98.5 \pm 1.4$ %, $PTV_{33Gy}$: $94.6 \pm 4.8$ %) and clinical plans ($PTV_{25Gy}$: $99.8 \pm 0.2$ %, $PTV_{33Gy}$: $94.7 \pm 1.2$ %). We observe smaller PTV coverage variations on the bot plans because the score function does not reward coverages beyond 95% coverage. [13] The mean MU value of the bot plans is higher ($1995 \pm 351$ MU) than that of clinical plans ($1742 \pm 271$ MU), indicating the complexity of the bot plan is slightly higher than that of the clinical plans.

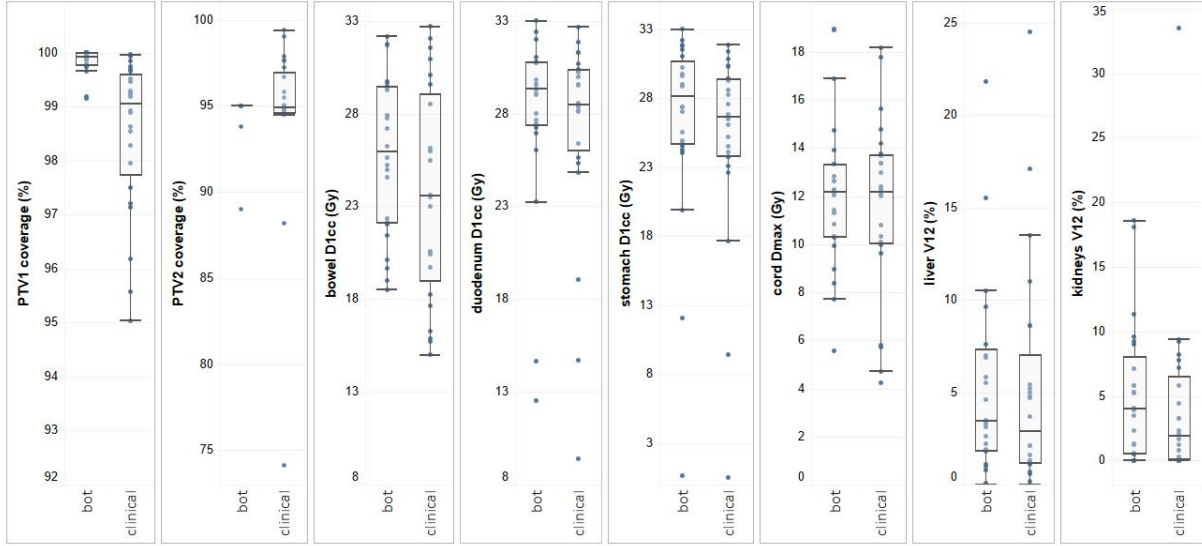

**Figure 2**. Dosimetric comparison between RL bot plans and clinical plans. The boxes represent quartiles, and the whiskers mark the datapoints within the 1.5 interquartile ranges (IQR) from the median values. The clinical constraints for bowel $D_{1cc}$, duodenum $D_{1cc}$, stomach $D_{1cc}$ are 33 Gy. Cord $D_{max}$ is limited below 20 Gy, and kidney $V_{12Gy}$ is limited below 25%-50%. All clinical plans and RL bot plans meet these clinical constraints.

Figure 3 shows dose distributions of two randomly selected validation plans (Fig. 3d-3f, Fig. 3j-3l) and their corresponding clinical plans (Fig. 3a-3c, Fig. 3g-3i). The RL-plans show similar PTV coverages compared to the clinical plans. However, the RL plans tend to exhibit better conformity on 33 Gy isodose lines but tend to over cover $PTV_{25Gy}$. This is likely due to the fact that the score function $S$ does not explicitly penalize dose spill out of the primary PTV into non-OAR regions.

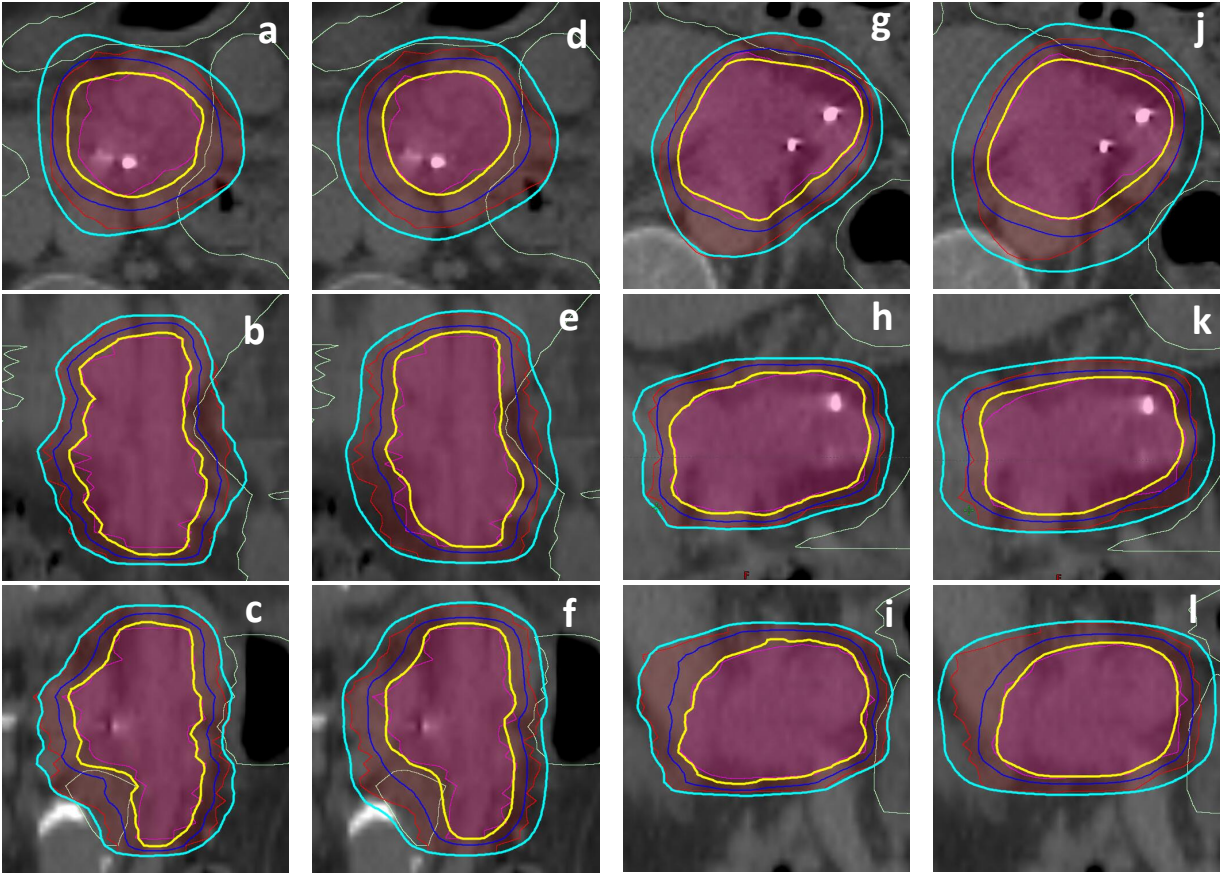

**Figure 3**. Cross-sections of two randomly selected clinical plans: (a-c) (g-i) and the corresponding RL plans (d-f) (j-l) . The three rows, from top to bottom, are axial, coronal, and sagittal views, respectively. The prescription doses to the primary PTV (red segments) and the boost PTV (magenta segments) are 25 Gy and 33 Gy, respectively. The 25 Gy and 33Gy isodose lines are represented by cyan and yellow lines. The dose limit to GI luminal structures (light green contours) is 33Gy less than 1cc.

### 3.2. Knowledge interpretability

The feature weighting factors $\theta^T$ contains information regarding the expected plan quality change, measured by the plan quality score function *S*, after a certain action at a certain state. An action is usually considered optimal when the feature value vector is well aligned with the corresponding row on $\theta^T$. This characteristic of the model makes the model readily interpretable. Figure 4 shows two regions of the reshaped $\theta^T$. The full feature map is shown in the supplementary material document.

Figure 4a illustrates that the bot has learned that when both $PTV_{33Gy}$ coverage and stomach $D_{1cc}$ constraints are compromised, it should consider adding lower constraint to an auxiliary structure that avoids the overlapping region between the PTV and the stomach. In contrast, it is often not effective to directly add PTV lower constraints. Similarly, Fig. 4b shows that adding stomach+6mm upper constraints is preferred when $PTV_{33Gy}$ $D_{98\%}$ is slightly violated and the stomach $D_{1cc}$ dose constraint is violated. Such learned knowledge is consistent with our planning experience. Therefore, we conclude that the RL bot learns to make sensible choices given the state information and our formulation of the action-value function offers meaningful insights into the learned planning strategies in the form of a "knowledge map". The RL learning provides a systematic and subjective methodology of learning planning knowledge.

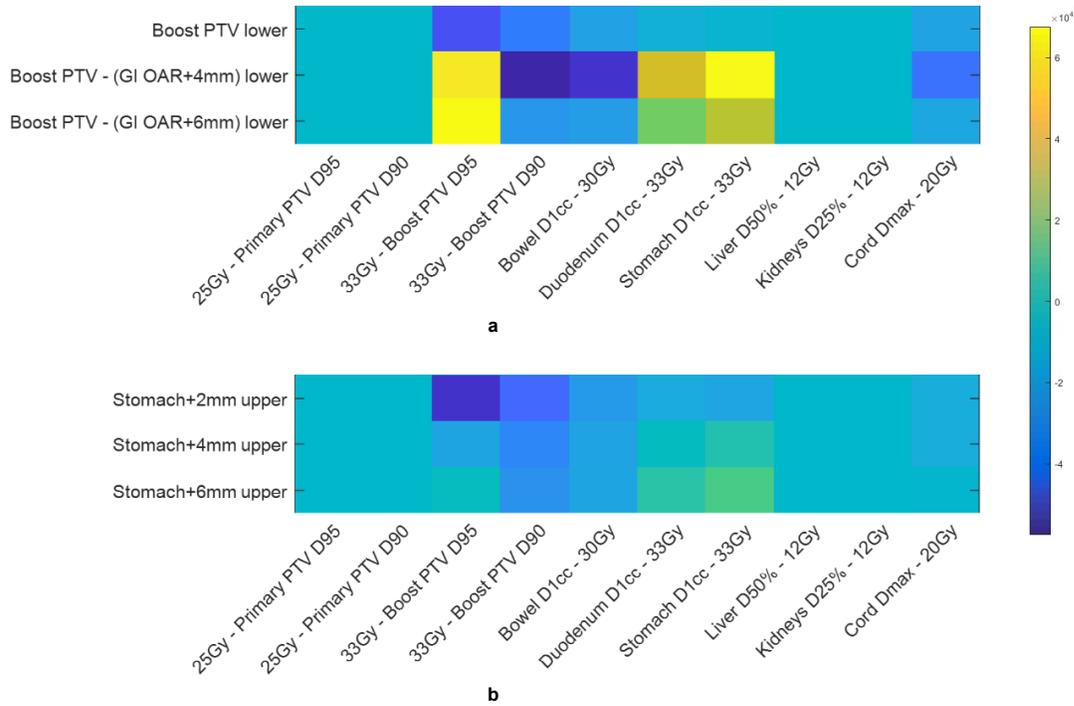

**Figure 4.** The weighting vector $\theta^T$ reshaped based on features and actions corresponding to: (a) PTV coverage and (b) stomach constraints. The weightings are of arbitrary units. At each bot-TPS interaction, we get action-value $Q(s,a)$ by multiplying $\theta^T$ by the feature vector $\varphi(s,a)$, which is evaluated in the TPS at the step.

### 3.3. Knowledge reproducibility

Our experiments have also demonstrated that the training of the RL bot is highly reproducible. Figure 5 shows the average differences of feature weighting factors $\theta^T$ learned in two separate training sessions. The average absolute change is 2.5%. Considering that the training sessions involve substantial introduced stochasticity, the differences between the two knowledge maps are relatively small, which preliminarily shows that the model training procedure is reproducible.

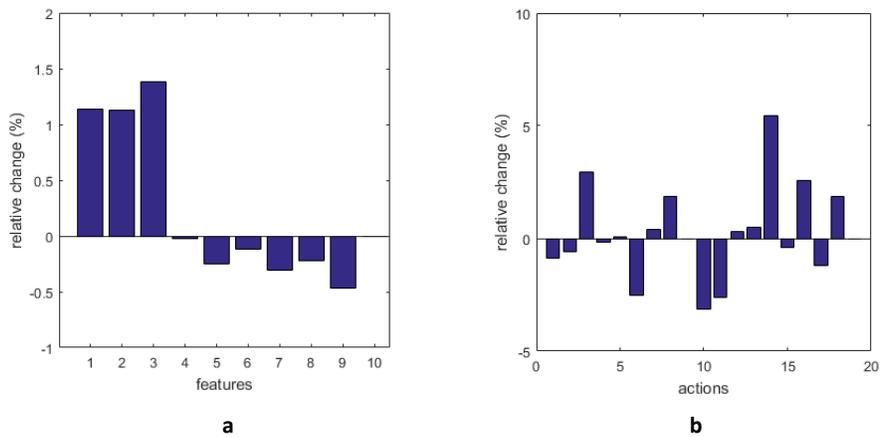

**Figure 5.** Average knowledge map differences across (a) different features and (b) different actions.

## 4. Discussion

In pancreas SBRT treatment planning, GI structures (small bowel including duodenum, large bowel and stomach) are often the structures limiting full boost PTV coverage, due to its proximity to the boost PTV. Planners iteratively evaluate the quality of boost PTV dose coverage with respect to the GI constraints and make adjustments accordingly. Notably, several actions are often taken when a planner modifies a plan, including adjusting priority or placement of existing structures and adding auxiliary structures to guide the local/regional dose dispositions, in both volume size and dose levels. We have formulated this process into a finite-horizon reinforcement-learning framework, the crucial components of which include states, actions, and rewards. First, we have discretized the states, in a similar fashion to how planners evaluate plans (i.e. constraint satisfaction). Second, we have identified a set of common actions that planners would take to address different planning issues, such as insufficient coverage, dose spill, etc. Third, we have derived a reward system based on our physicians' input. Finally, we have managed to limit the complexity of the system and thereby created a planning bot that can be implemented in a clinical TPS.

The training stage of the planning bot essentially simulates the learning process of a human planner. The bot first takes many attempts in trying different actions at different states, and after each action, the plan is re-evaluated, and a reward is assigned accordingly. As the bot gains planning experience gradually, it makes decisions with the guidance of retained prior knowledge, but also makes attempts to explore the alternative methods for surprise gains. After completing the training process, the bot has acquired knowledge that can guide it to get the highest plan quality possible. The knowledge, summarized in an action value function, contains the information of expected long-term rewards of taking certain actions at certain states. When planning a new patient case, the bot periodically evaluates the current state of the plan, infers the best option from the action value function, and takes the corresponding action, thus completing the navigation of autonomous planning process. To fully utilize the geometrical information contained in the training dataset, we have augmented the training dataset by expanding and shrinking the boost PTVs. This step effectively allows the bot to practice planning on sufficient anatomical variations without requiring more training cases. We have introduced variations on the boost PTV because the primary focus of the planning bot is to effectively handle the contradicting boost PTV coverage requirements and GI OAR 1cc constraints. Similar augmentation methods can potentially be applied to increase the variations on other OARs. During the development of the planning bot, we tuned the RL model by using the plan quality scores of a few hold-out training cases to gauge the performance of the bot. Specifically, we have determined the number of actions necessary and the number of cases required for model training, in addition to the model parameters such as $\varepsilon$ and $N$. We have estimated that more than 10-20 cases are necessary to train the bot, though the number of cases needed is dependent on the degree of anatomy variations for the treatment site and the requirements of the planning task.

The limitations of this model are twofold. First, the linear approximation used in this work, while interpretable, potentially limits the flexibility of the model when approximating the underlying action values, i.e. the ground truth of the expected long-term reward, for more complicated planning tasks. In this study, we used a SARSA algorithm with linear action value function approximation to determine optimal actions. It may be necessary to investigate the use of other types of value function classes, such as deep neural networks [11]. However, a more complicated model is expected to have less interpretability and require more tuning, both of which are undesirable for clinical applications. Another potential limitation of the proposed method is that the actions have to be discretized to fit in the SARSA framework. This is reflected in the fact that we have set the planning constraint priorities to be constant. The priorities were selected carefully such that placing a constraint introduces sufficient plan changes and yet does not overshoot. With policy gradient-based RL algorithms, it is possible to learn the optimal policy directly in a continuous domain. This class of algorithms is also worth investigating for performing treatment planning tasks. In this study, we have applied the RL planning bot to solving a challenging planning task known to be heavily reliant on planner input. However, the RL bot is not limited to this specific treatment planning task. The model can be adopted for other treatment sites by using a different set of features and actions to match the planning practices. In addition, the plan quality score should be re-defined to reflect the clinical plan quality preferences.

Previously, automated planning based on supervised machine learning has gained widespread acceptance in the radiation therapy community[13-16] and has been implemented in commercial TPS[17]. This class of algorithms, collectively referred to as knowledge-based planning (KBP), train a model to represent the correlation between patient anatomy and dose distribution based on previously treated patients. For a new patient, KBP predicts the best achievable OAR DVHs and generates corresponding dose-volume constraints as input for plan optimization. Compared with KBP, the RL bot is different in two aspects. Firstly, the bot does not rely on optimal plans in training data. The underlying assumption of KBP is that the plans used for model training are optimal under the current standard. In contrast, the RL bot acquires planning knowledge by trial-and-error and thereby does not require previous plans. As a result, when a planning protocol gets updated, the planning bot can be simply re-trained with updated score function while KBP cannot be used until enough new plans have been collected and the model can be re-trained. Secondly and more importantly, KBP places a set of estimated dose-volume constraints for optimization. This method, while performing well for many treatment sites, is not sufficient to address the complex local tradeoffs in pancreas SBRT. The lack of spatial information in dose-volume constraints results in inefficient cost function assignment and the planner often needs to create local optimization structures to encode the spatial information manually, which is time-consuming and defeats the purpose of auto-planning.

To our knowledge, there have been very few publications applying reinforcement learning to external beam treatment planning tasks, and this is the first work on implementing RL planning in a clinical TPS. Shen *el al*. has recently proposed a deep reinforcement learning-based prostate IMRT virtual planner, which utilizes neural networks to adjust dose-volume constraints [18]. They have shown that the virtual planner improves plan quality upon the initialized plans and is a potentially promising planning method, acknowledging that the method can be a "black box". In contrast, we have made significant efforts to simplify the model to improve transparency and demonstrate the efficacy of an autonomous, yet interpretable planning bot powered by reinforcement learning. We narrowed down features to a limited set of variables summarized from domain knowledge, namely those commonly used by our planners to examine the treatment plans before implementing manual plan changes. Also, we used linear function approximation for the action value determination, which presents a simple and interpretable model. In this study, we have focused on the planning of pancreas SBRT treatments. However, the proposed framework should apply to other treatment sites with careful design of features and actions.

## 5. Conclusion

The planning bot generates clinically acceptable plans by taking consistent and predictable actions. Additionally, the knowledge maps learned in separate training sessions are consistent, and the knowledge learned by the RL bot is consistent with human planning knowledge. Therefore, the training phase of our planning bot is tractable and reproducible, and the knowledge obtained by the bot is interpretable. As a result, the trained planning bot can be validated by human planners and serve as a robust planning assistance routine in the clinics.